\documentclass[12pt,preprint2]{aastex}

\usepackage{lscape}

\shortauthors{Onken \& Kollmeier}
\shorttitle{\ion{Mg}{2}-Based BH Masses}

\begin{document}

\title{An Improved Method for Using \ion{Mg}{2} to Estimate Black Hole
Masses in Active Galactic Nuclei} 

\author{Christopher A.\ Onken\altaffilmark{1,2} \& Juna A.\
Kollmeier\altaffilmark{3}}

\altaffiltext{1}{Dominion Astrophysical Observatory, Herzberg Institute of
  Astrophysics, National Research Council of Canada, 5071 West Saanich Road,
  Victoria, BC, V9E 2E7, Canada}

\altaffiltext{2}{Current address: Research School of Astronomy and
  Astrophysics, The Australian National University, Cotter Road, Weston Creek,
  ACT 2611, Australia}

\altaffiltext{3}{Observatories of the Carnegie Institution of Washington, 813
  Santa Barbara Street, Pasadena, CA 91101}

\email{onken@mso.anu.edu.au, jak@ociw.edu}

\begin{abstract}

  We present a method for obtaining accurate black hole (BH) mass estimates
  from the \ion{Mg}{2} emission line in active galactic nuclei (AGNs).
  Employing the large database of AGN measurements from the Sloan Digital Sky
  Survey (SDSS) presented by Shen et al., we find that AGNs in the redshift
  range 0.3--0.9, for which a given object can have both H$\beta$\ and
  \ion{Mg}{2} line widths measured, display a modest but correctable
  discrepancy in \ion{Mg}{2}-based masses that correlates with the Eddington
  ratio. We use the SDSS database to estimate the probability distribution of
  the true (i.e., H$\beta$-based) mass given a measured \ion{Mg}{2} line
  width. These probability distributions are then applied to the SDSS
  measurements from Shen et al.\ across the entire \ion{Mg}{2}-accessible
  redshift range (0.3--2.2).  We find that accounting for this residual
  correlation generally increases the dispersion of Eddington ratios by a
  small factor ($\sim$0.09~dex for the redshift and luminosity bins we
  consider). We continue to find that the intrinsic distribution of Eddington
  ratios for luminous AGNs is extremely narrow, 0.3--0.4~dex, as demonstrated
  by Kollmeier et al.  Using the method we describe, \ion{Mg}{2} emission
  lines can be used with confidence to obtain BH mass estimates.

\end{abstract}


%\clearpage

\section{INTRODUCTION}

The $M_{\rm BH}$-$\sigma$ correlation between supermassive black hole (BH)
mass and the velocity dispersion of the surrounding stellar system indicates a
significant connection between galaxy and BH assembly \citep{ferrarese00,
  gebhardt00}.  Thus, it is essential to establish the most basic BH
parameters: their intrinsic distribution of masses and growth rates.

It is extraordinarily difficult to measure BH masses directly, even in the
nearby universe, because of the small spatial scales that must be resolved to
probe the gravitational influence of the BH. In active galactic nuclei (AGNs),
the technique of reverberation mapping \citep{blandford82,peterson93} employs
high resolution in the time domain to probe gas dynamics on spatial scales
close to the BH. However, the long-duration spectroscopic monitoring campaigns
required for reverberation studies currently preclude the method from being
applied to large numbers of objects. Therefore, one must rely on even more
indirect techniques of BH mass estimation to build up statistically
significant samples.

The ``virial'' method has been empirically calibrated from reverberation
mapping experiments \citep{wandel99, vestergaard02, mclure02} and allows a BH
mass estimate from a one-time measurement of the width of a broad emission
line and the AGN luminosity (see \S~2). To facilitate application of the
virial technique to large optical AGN surveys, versions have been developed
for H$\beta$\ at low redshift, for the \ion{Mg}{2} doublet near 2800\AA\ at
intermediate redshift, and for the \ion{C}{4} doublet near 1550\AA\ at high
redshift \cite[for recent prescriptions, see][]{vestergaard06,mcgill08}.

In this paper, we present evidence of a systematic discrepancy in
\ion{Mg}{2}-based BH mass estimates as a function of Eddington ratio (the
ratio of the bolometric luminosity, $L_{\rm bol}$, to the luminosity required
for radiation pressure to balance the gravity of the BH), as well as a method
to correct this trend.

\section{Method of Analysis \label{sec:method}}

The standard equation for estimating BH masses in AGNs from single-epoch
spectroscopy is:
\begin{equation}
\log M_{\rm BH} = a + b \log L_c + 2 \log V,
\end{equation}
where $V$ is the width of the broad emission line, $L_c$ is the continuum
luminosity near the line, and $a$ and $b$ are constants (which vary from line
to line).  The dependence on $L_c$ arises because the distance of the broad
emission line gas from the BH has been observed to correlate tightly with the
AGN luminosity over 4 orders of magnitude in $L_c$ \citep{bentz07}.  Thus, the
mass equation reverts to the simple, virial combination of radius and
velocity. The calibration of these relations rests on the bedrock of
reverberation mapping measurements of local AGNs. By far, the best
reverberation mapping dataset exists for H$\beta$\ \citep{peterson04}. The $a$
and $b$ coefficients for both \ion{C}{4} and \ion{Mg}{2} principally rely on
empirical correlations with H$\beta$-reverberation masses. However, whereas
\ion{C}{4} reverberation studies of a handful of objects are consistent with
expectations from H$\beta$\ \citep{peterson05}, a clear indication of
reverberation has not yet been found for \ion{Mg}{2}. Although the line has
been seen to vary in both flux and width
\cite[e.g.,][]{clavel91,dietrich95,metzroth06,woo08}, only weak \ion{Mg}{2}
reverberation signals have been seen \citep{reichert94}.  Thus, the mass
relation for \ion{Mg}{2} relies on the correlation of single-epoch estimates
with H$\beta$\ masses and the argument that \ion{Mg}{2} and H$\beta$\ have
similar ionization potentials \citep{mclure04}.

With the wide wavelength coverage of Sloan Digital Sky Survey (SDSS) spectra,
certain redshift windows allow for two of these three emission lines to be
measured simultaneously: H$\beta$\ and \ion{Mg}{2} are both accessible for
$z\sim$0.3--0.9, while \ion{Mg}{2} and \ion{C}{4} can both be measured for
$z\sim$1.7--2.2. By comparing the BH mass estimates from the two lines, we are
thus able to study systematic trends. Such a comparison has been done
previously for the {\it mean relation}, but the large sample size of the SDSS
permits an analysis of higher-order correlations, which prove to be quite
important.

Shen et al.\ (2008) have provided line width (FWHM) measurements and BH mass
estimates---using the relations of \citet{mclure04} and Vestergaard \& Peterson (2006)---for roughly 60,000 AGNs from the SDSS, including
$\sim$8,000 with both H$\beta$\ and \ion{Mg}{2}, and $\sim$15,000 with both
\ion{Mg}{2} and \ion{C}{4}. They provide a detailed analysis of the
relationship between \ion{Mg}{2} and \ion{C}{4}, but simply describe the ratio
of H$\beta$\ to \ion{Mg}{2} FWHM as following a log-normal distribution with a
mean of 0.0062~dex and a dispersion of 0.11~dex. Shen et al.\ noted that the
relation of the two FWHMs deviates slightly from a perfect correlation, but
did not explore the issue further.

\begin{figure}
\plotone{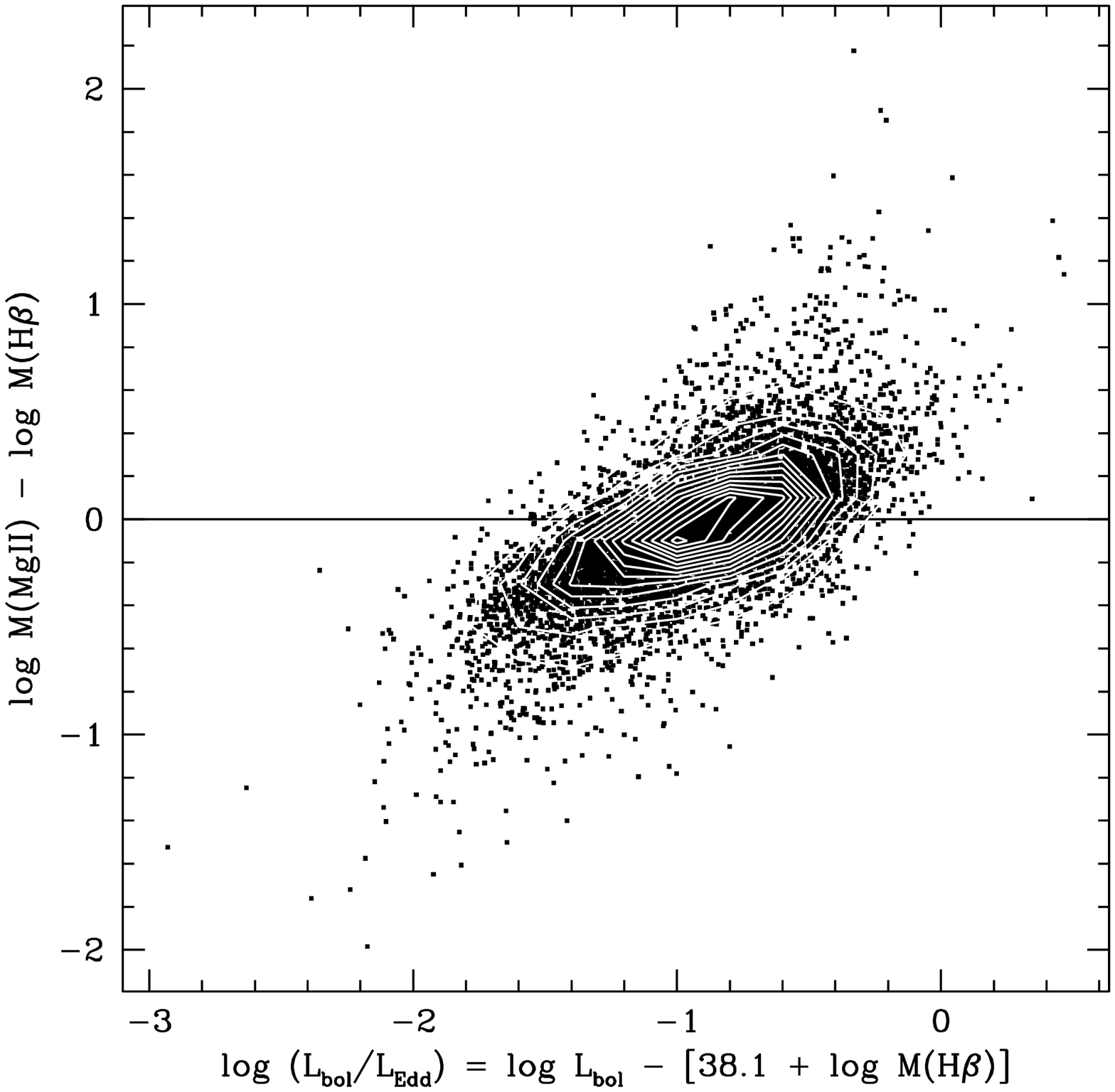}
\caption{Difference in H$\beta$\ and \ion{Mg}{2} estimates of $\log M_{\rm
    BH}$ as a function of the log of the Eddington ratio. The Eddington ratio
  is calculated using the H$\beta$\ mass estimate. White contours are linearly
  spaced at intervals of 50 AGNs per 0.2$\times$0.2-dex bin. The BH mass
  estimates from \ion{Mg}{2} are correlated with the Eddington ratio such that
  the mass is underpredicted at low Eddington ratio and overpredicted at high
  Eddington ratio.}
\label{fig:mass}
\end{figure}

Under the premise that H$\beta$, being the most extensively
reverberation-mapped emission line, provides the best indicator of the BH
mass, we examine AGNs having both H$\beta$\ and \ion{Mg}{2} mass
measurements. Due to uncertainties in the line width measurements, we exclude
the small number of AGNs that were flagged by Shen et al.\ as broad absorption
line objects. However, the inclusion of these objects has negligible effect on
our results.  In Figure~1, we plot the difference in the (log of the) BH mass
as a function of Eddington ratio\footnote{The Eddington ratio is computed with
  the H$\beta$-based BH mass.}. The strong correlation implies that if
\ion{Mg}{2} is calibrated simply from the mean of the H$\beta$-\ion{Mg}{2}
relation, then it will underestimate the BH mass at low Eddington ratio and
overestimate the mass at high Eddington ratio\footnote{We note that the same
  trend from Figure~\ref{fig:mass} is seen when replacing FWHM with the
  inter-percentile value measurements of Fine et al.\ (2008).}.  This would
lead one to infer a narrower distribution of Eddington ratios than actually
exists.  One way to quantify this effect is to measure the slope of the
correlation

\begin{equation}
\beta = \tan \biggl[{1\over 2}\arctan{2 c_{12}\over c_{11} - c_{22}}\biggr]
\end{equation}

\noindent where $c_{ij}$ is the covariance matrix of the distribution shown in
Figure~\ref{fig:mass}.  A slope $\beta=0$\ would imply that \ion{Mg}{2}
provided completely independent information on the BH mass, while $\beta=1$\
would imply that no information is conveyed by the \ion{Mg}{2} measurement.
The actual value is $\beta=0.76$, which means that \ion{Mg}{2} {\it is}
indicative of the BH mass, but must be treated with care.

Simply modifying the luminosity-dependence of the \ion{Mg}{2} mass formula
(i.e., changing $b$ in eq.~[1]) cannot remove the observed trend. Therefore,
to make a statistical correction to the \ion{Mg}{2}-based masses, we adopt the
following method. For the SDSS objects with both H$\beta$\ and \ion{Mg}{2}
FWHMs, we look at AGNs with \ion{Mg}{2} lines in a given 0.1~dex bin of FWHM
and tabulate the distribution of H$\beta$\ FWHMs for those objects. We take
those H$\beta$\ distributions (normalized appropriately) as the probability
distributions for the true FWHM underlying the observed \ion{Mg}{2} value
(Fig.~2; Table~1).

For any object with an accessible \ion{Mg}{2} line, each bin in true FWHM is
combined with the observed continuum luminosity to calculate a BH mass (via
eq.~[1]), and the mass is then used with the object's bolometric luminosity to
derive an Eddington ratio. The probability for each bin of FWHM is added to
the total number of objects contributing to the corresponding Eddington ratio
bin (which are 0.2~dex-wide, allowing a one-to-one match between FWHM and
Eddington ratio bins). Thus, each \ion{Mg}{2} FWHM becomes a weighted
distribution of Eddington ratios\footnote{If the observed \ion{Mg}{2} FWHM
  falls in a bin in which there were no H$\beta$+\ion{Mg}{2} measurements, it
  is given a probability of 1 within the bin corresponding to the \ion{Mg}{2}
  FWHM.  In the sample we consider, this applies to a single AGN.}, while
H$\beta$- and \ion{C}{4}-based Eddington ratios contribute directly at their
observed values.

We now apply this technique to the Shen et al.\ measurements of the uniformly
selected subsample of SDSS AGNs (Richards et al.\ 2006).

\section{Results \& Discussion \label{sec:results}}

We construct distributions of Eddington ratios for SDSS in ranges of ($L_{\rm
  bol}$, $z$) in Figure~3, showing the results both with (solid) and without
(dotted) the \ion{Mg}{2} substitution presented above. The statistics of the
distributions are given in Table~2. After correcting the \ion{Mg}{2}
measurements in the uniformly selected SDSS sample, we find that the average
width of the Eddington ratio distribution increased by 0.09~dex for objects in
the redshift range 0.3--2.2. We therefore find that the distribution of
Eddington ratios remains very narrow at $\sim$0.4~dex, as was found by
\citet{kollmeier06} for the AGES-I survey.

To test our procedure, we apply it to multiple subsamples for which we have
both H$\beta$\ and \ion{Mg}{2} data.  Figure~\ref{fig:check} shows the
distribution of Eddington ratios for each subsample calculated in two ways,
first using the true H$\beta$\ mass (solid) and second using our procedure
applied to the \ion{Mg}{2} mass (dashed).  The similarity of these
distributions (and the differences from the raw \ion{Mg}{2}-based values,
shown as the dotted histograms) demonstrates that our procedure works well,
recovering the true Eddington ratio distribution from the \ion{Mg}{2} derived
masses.

The general trend we observe could be explained physically if the
location where lines are formed in the broad-line region depends on
accretion rate. In this case, the radius-luminosity relation would
also depend on Eddington ratio and that would introduce an additional
term in the virial mass relation.  It would then be possible, in
principle, to remove this dependence entirely analytically by fitting
the observed correlation.  We investigate this further in an upcoming
work.

The virial method for BH mass estimation has opened a new window in the study
of supermassive BH demographics.  While each indicator has systematics that
must be addressed (i.e., asymmetric line profiles, contamination from disk
winds and metal lines, etc.), it is important to understand and attempt to
correct for these systematics so that this technique can be applied with
confidence.  In this contribution, we have identified a limitation in
estimating BH masses from the \ion{Mg}{2} line and presented a way to remove
it by exploiting the overlap of \ion{Mg}{2} and H$\beta$\ measurements
presented by Shen et al.\ (2008).  This method puts \ion{Mg}{2} masses more
securely on the same scale as H$\beta$, which should be the most reliable in
these studies.  As the bias in the \ion{Mg}{2} masses is also seen in our
analysis of the 2dF Quasar Redshift Survey (2QZ) and the AGN and Galaxy
Evolution Survey (AGES), in forthcoming work, we will apply our correction 
technique to those datasets (including the expanded sample of AGES-II).

Understanding the transition between high-luminosity, high-redshift AGNs
having a narrow Eddington ratio distribution ($\sim$0.4~dex) and
low-luminosity, low-redshift AGNs with a broad Eddington ratio distribution
($>$1~dex; Ho 2002; Woo \& Urry 2002) can provide important constraints on the
physics of BH accretion. To determine these distributions, it is critical to
continue to improve BH mass estimates that can be used through the bulk of the
cosmic AGN activity.

\acknowledgments We thank the anonymous referee for providing useful
feedback. We thank Andy Gould for helpful discussions and a careful reading of
an early draft of this letter. We thank Chris Kochanek for insightful comments
on this manuscript. We thank Matthias Dietrich for valuable suggestions, and
Pat Hall and Alireza Rafiee for stimulating discussions. CAO acknowledges
support by a Plaskett Fellowship during the completion of this work. JAK
acknowledges the support of Hubble Fellowship HF-01197 and a
Carnegie-Princeton fellowship during the completion of this project. This
research was supported in part by the National Science Foundation under Grant
No. PHY05-51164. Funding for the SDSS and SDSS-II has been provided by the
Alfred P. Sloan Foundation, the Participating Institutions, the National
Science Foundation, the U.S. Department of Energy, the National Aeronautics
and Space Administration, the Japanese Monbukagakusho, the Max Planck Society,
and the Higher Education Funding Council for England. The SDSS Web Site is
http://www.sdss.org/.

%The SDSS is managed by the Astrophysical Research Consortium for the
%Participating Institutions. The Participating Institutions are the American
%Museum of Natural History, Astrophysical Institute Potsdam, University of
%Basel, University of Cambridge, Case Western Reserve University, University of
%Chicago, Drexel University, Fermilab, the Institute for Advanced Study, the
%Japan Participation Group, Johns Hopkins University, the Joint Institute for
%Nuclear Astrophysics, the Kavli Institute for Particle Astrophysics and
%Cosmology, the Korean Scientist Group, the Chinese Academy of Sciences
%(LAMOST), Los Alamos National Laboratory, the Max-Planck-Institute for
%Astronomy (MPIA), the Max-Planck-Institute for Astrophysics (MPA), New Mexico
%State University, Ohio State University, University of Pittsburgh, University
%of Portsmouth, Princeton University, the United States Naval Observatory, and
%the University of Washington.

%\clearpage

%%figures

\clearpage

\thispagestyle{empty}

\begin{landscape}
\begin{deluxetable}{clllllllllllllll}
\tabletypesize{\tiny}
\tablecaption{\ion{Mg}{2} Replacement Matrix}
\tablewidth{0pt}
\tablehead{
\colhead{} & \multicolumn{15}{c}{Probability} \\
\cline{2-16}\\
\colhead{MgII FWHM} & \colhead{3.05} & \colhead{3.15} & \colhead{3.25} & \colhead{3.35} & \colhead{3.45} & \colhead{3.55} & \colhead{3.65} & \colhead{3.75} & \colhead{3.85} & \colhead{3.95} & \colhead{4.05} & \colhead{4.15} & \colhead{4.25} & \colhead{4.35} & \colhead{4.45}
}
\startdata
3.05 & 1.0000 & 0.0000 & 0.0000 & 0.0000 & 0.0000 & 0.0000 & 0.0000 & 0.0000 & 0.0000 & 0.0000 & 0.0000 & 0.0000 & 0.0000 & 0.0000 & 0.0000\\
3.15 & 0.0000 & 1.0000 & 0.0000 & 0.0000 & 0.0000 & 0.0000 & 0.0000 & 0.0000 & 0.0000 & 0.0000 & 0.0000 & 0.0000 & 0.0000 & 0.0000 & 0.0000\\
3.25 & 0.0000 & 0.0000 & 1.0000 & 0.0000 & 0.0000 & 0.0000 & 0.0000 & 0.0000 & 0.0000 & 0.0000 & 0.0000 & 0.0000 & 0.0000 & 0.0000 & 0.0000\\
3.35 & 0.0000 & 0.0361 & 0.0723 & 0.0964 & 0.3494 & 0.2410 & 0.0602 & 0.0602 & 0.0482 & 0.0000 & 0.0120 & 0.0120 & 0.0120 & 0.0000 & 0.0000\\
3.45 & 0.0000 & 0.0038 & 0.0379 & 0.1061 & 0.2386 & 0.3182 & 0.1667 & 0.0530 & 0.0303 & 0.0227 & 0.0152 & 0.0076 & 0.0000 & 0.0000 & 0.0000\\
3.55 & 0.0000 & 0.0009 & 0.0196 & 0.0559 & 0.1985 & 0.3150 & 0.2591 & 0.0801 & 0.0270 & 0.0261 & 0.0121 & 0.0028 & 0.0009 & 0.0000 & 0.0019\\
3.65 & 0.0000 & 0.0000 & 0.0050 & 0.0153 & 0.0761 & 0.2413 & 0.3625 & 0.2061 & 0.0532 & 0.0264 & 0.0096 & 0.0023 & 0.0023 & 0.0000 & 0.0000\\
3.75 & 0.0000 & 0.0008 & 0.0021 & 0.0076 & 0.0239 & 0.0877 & 0.2244 & 0.3410 & 0.2072 & 0.0751 & 0.0227 & 0.0046 & 0.0021 & 0.0008 & 0.0000\\
3.85 & 0.0000 & 0.0007 & 0.0022 & 0.0126 & 0.0119 & 0.0491 & 0.0908 & 0.1555 & 0.2924 & 0.2619 & 0.1057 & 0.0134 & 0.0037 & 0.0000 & 0.0000\\
3.95 & 0.0000 & 0.0000 & 0.0000 & 0.0000 & 0.0000 & 0.0000 & 0.0000 & 0.0000 & 0.0000 & 1.0000 & 0.0000 & 0.0000 & 0.0000 & 0.0000 & 0.0000\\
4.05 & 0.0000 & 0.0000 & 0.0000 & 0.0412 & 0.0619 & 0.1031 & 0.1340 & 0.1340 & 0.0722 & 0.1237 & 0.1856 & 0.0825 & 0.0309 & 0.0309 & 0.0000\\
4.15 & 0.0000 & 0.0000 & 0.0000 & 0.0909 & 0.1818 & 0.0455 & 0.0909 & 0.0455 & 0.1364 & 0.0909 & 0.1364 & 0.0455 & 0.1364 & 0.0000 & 0.0000\\
4.25 & 0.0000 & 0.0000 & 0.3333 & 0.0000 & 0.0000 & 0.3333 & 0.3333 & 0.0000 & 0.0000 & 0.0000 & 0.0000 & 0.0000 & 0.0000 & 0.0000 & 0.0000\\
4.35 & 0.0000 & 0.0000 & 0.4000 & 0.0000 & 0.0000 & 0.0000 & 0.2000 & 0.2000 & 0.0000 & 0.0000 & 0.0000 & 0.0000 & 0.0000 & 0.2000 & 0.0000\\
4.45 & 0.0000 & 0.0000 & 0.0000 & 0.0000 & 0.0000 & 0.0000 & 0.0000 & 0.0000 & 0.0000 & 0.0000 & 0.0000 & 0.0000 & 0.0000 & 0.0000 & 1.0000\\
\enddata
\tablecomments{The values listed are the probabilities of the H$\beta$ FWHM
  falling within a particular bin, given an input MgII FWHM. All
  velocities are in $\log$ km~s$^{-1}$, and all bins are $\pm$0.05~dex.}
\end{deluxetable}
\end{landscape}

\clearpage

\begin{deluxetable}{lllccccccccc}
\tabletypesize{\footnotesize}
\tablecaption{Statistics of Eddington Ratio Distributions for SDSS}
\tablewidth{0pt}
\tablehead{
\colhead{} & \colhead{} & \colhead{} & \multicolumn{2}{l}{Raw} &
\colhead{} & \colhead{} & \colhead{} & \multicolumn{4}{l}{\ion{Mg}{2} Replaced} \\
 \cline{4-7} \cline{9-12} \\
\colhead{$z_{\rm bin}$} & \colhead{$\log L_{\rm bin}$} & \colhead{$N$} &
\colhead{$\mu$} & \colhead{$\sigma$} & \colhead{$S_k$} & \colhead{$A_k$} &
\colhead{} & \colhead{$\mu$} & \colhead{$\sigma$} & \colhead{$S_k$} & \colhead{$A_k$}
}
\startdata
0.3--1.2 & $<46$ & 6187 & $-0.97$ & 0.33 & $-0.48$ & 4.27 & & $-0.98$ & 0.39 & $-0.33$ & 3.30 \\
1.2--2.2 & $<46$ & 15 & $-1.03$ & 0.40 & \hphantom{$-$}0.55 & 2.07 & & $-0.92$ & 0.44 & \hphantom{$-$}0.05 & 2.17 \\
0.3--1.2 & 46--46.5 & 5233 & $-0.77$ & 0.25 & $-0.24$ & 3.39 & & $-0.79$ & 0.36 & $-0.04$ & 2.95 \\
1.2--2.2 & 46--46.5 & 5360 & $-0.76$ & 0.24 & $-0.02$ & 3.51 & & $-0.72$ & 0.35 & \hphantom{$-$}0.05 & 2.82 \\
0.3--1.2 & $>46.5$ & 789 & $-0.61$ & 0.22 & $-0.47$ & 3.17 & & $-0.61$ & 0.34 & $-0.06$ & 2.92 \\
1.2--2.2 & $>46.5$ & 9860 & $-0.65$ & 0.25 & \hphantom{$-$}0.06 & 3.73 & & $-0.65$ & 0.34 & \hphantom{$-$}0.24 & 3.27 \\
\enddata

\tablecomments{For each bin in redshift $z_{\rm bin}$ and luminosity $L_{\rm
    bin}$, the table lists: $N$, the number of AGNs; $\mu$, the mean logarithm
  of the Eddington ratio; $\sigma$, the dispersion in $\log(L_{\rm bol}/L_{\rm
    Edd})$; $S_k$, the skewness; and $A_k$, the kurtosis.}

\end{deluxetable}

\clearpage

\begin{figure}
\plotone{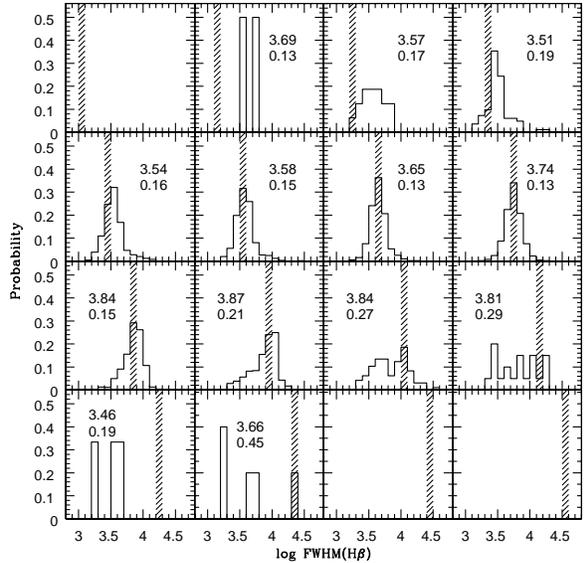}
\caption{Distribution of H$\beta$\ FWHMs for objects having \ion{Mg}{2} FWHM
  within the 0.1~dex-wide band shown as the cross-hatched bar in each
  panel. The mean and standard deviation in $\log$ FWHM(H$\beta$) is also
  indicated for each subsample. Despite the overall H$\beta$\ and \ion{Mg}{2}
  FWHM distributions having similar mean values, the total dispersion in log
  FWHM is 0.18~dex for the former and only 0.12~dex for the latter,
  demonstrating that \ion{Mg}{2} fails to reflect the full range of line
  widths (and hence BH masses).}
\label{fig:fwhm}
\end{figure}

\begin{figure}
\plotone{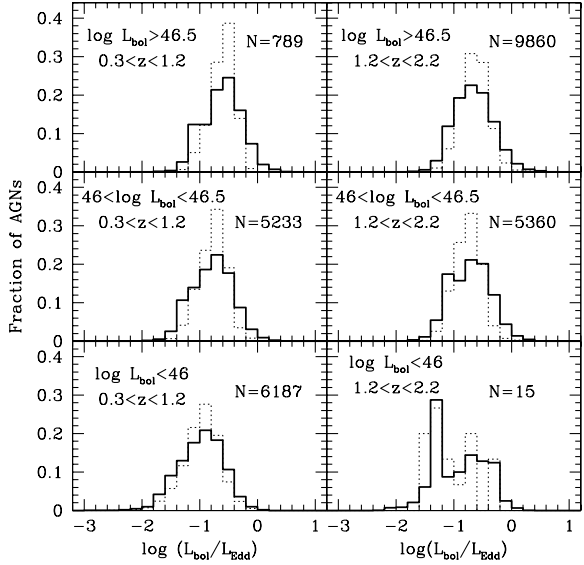}
\caption{Distribution of Eddington ratios in bins of bolometric luminosity and
  redshift, for the redshift range in which \ion{Mg}{2} estimates of the BH
  mass contribute to the distribution. Dotted histograms employ the raw mass
  estimates from all three emission lines. Solid histograms show the result
  after replacing each \ion{Mg}{2}-based mass with the corresponding
  probability distribution we derived from H$\beta$.  Bolometric luminosity,
  redshift, and the number of AGNs are shown in each panel.}
\label{fig:edd2}
\end{figure}

\begin{figure}
\plotone{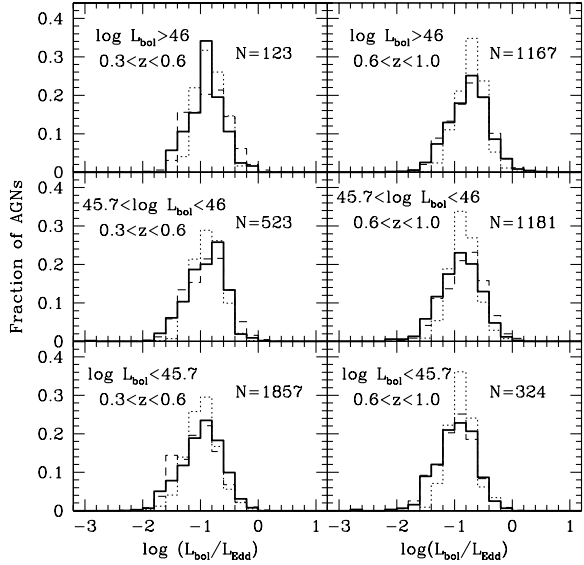}
\caption{Distribution of Eddington ratios in bins of bolometric luminosity and
  redshift, for objects where both H$\beta$\ and \ion{Mg}{2} masses were
  available.  Solid histograms show the distributions for the H$\beta$-derived
  masses, dashed histograms show the resultant distributions when our
  procedure is applied to the \ion{Mg}{2} masses, and the dotted histograms
  show the raw \ion{Mg}{2} estimates.  The similarity of the H$\beta$ (solid)
  and corrected (dashed) distributions demonstrates that our method of
  correcting the \ion{Mg}{2} masses correctly recovers the true H$\beta$\
  Eddington ratio distribution.  Bolometric luminosity, redshift, and the
  number of AGNs are shown in each panel.}
\label{fig:check}
\end{figure}

\end{document}